\documentclass[fleqn,usenatbib]{mnras}

\usepackage[T1]{fontenc}
\usepackage{ae,aecompl}
\usepackage{graphicx}    
\usepackage{amsmath}    
\usepackage{amssymb}    
\title[Radio observations of HD~41004]{In search of radio emission from exoplanets: GMRT observations of the binary system HD~41004}

\author[Mayank Narang]{
Mayank Narang$^{1}$,\thanks{E-mail: mayank.narang@tifr.res.in}
P. Manoj$^{1},$
C. H. Ishwara Chandra$^{2}$, Joseph Lazio$^3$, \newauthor
Thomas Henning$^4$, Motohide Tamura$^{5,6,7}$, Blesson Mathew$^{8}$, \newauthor Nitish Ujwal$^9$,  Pritha Mandal $^{8}$
\\
$^{1}$Tata Institute of Fundamental Research 
Homi Bhabha Road, Colaba, Mumbai 400005, India\\
$^{2}$National Centre for Radio Astrophysics, TIFR, Post Bag 3, Ganeshkhind, Pune 411007, India\\
$^{3}$ Jet Propulsion Laboratory, California Institute of Technology, Pasadena CA 91106,United States\\
$^4$ Max-Planck-Institut f\"{u}r Astronomie, K\"{o}nigstuhl 17, D-69117 Heidelberg, Germany\\
$^5$ The University of Tokyo, 7-3-1, Hongo, Bunkyo-ku, Tokyo, 113-0033, Japan\\
$^6$ Astrobiology Center, 2-21-1, Osawa, Mitaka, Tokyo 181-8588, Japan\\
$^7$ National Astronomical Observatory of Japan, 2-21-1 Osawa, Mitaka, Tokyo 181-8588, Japan\\
$^{8}$Department of Physics and Electronics, CHRIST (Deemed to be University), Bangalore, Karnataka - 560034\\
$^{9}$ Indian Institute of Technology Bombay, Powai, Mumbai, 400076, India 
}

\begin{document}
\label{firstpage}
\pagerange{\pageref{firstpage}--\pageref{lastpage}}
\maketitle

\begin{abstract}
This paper reports Giant Metrewave Radio Telescope (GMRT) observations of the binary system HD 41004 that are among the deepest images ever obtained at 150~MHz and 400 MHz in the search for radio emission from exoplanets. The HD 41004 binary system consists of a K1 V primary star and an M2 V secondary; both stars are host to a massive planet or brown dwarf. Analogous to planets in our solar system that emit at radio wavelengths due to their strong magnetic fields, one or both of the planet or brown dwarf in the HD 41004 binary system are also thought to be sources of radio emission. Various models predict HD~41004Bb to have one of the largest expected flux densities at 150 MHz. The observations at 150 MHz cover almost the entire orbital period of HD 41004Bb, and about $20\%$ of the orbit is covered at 400 MHz. We do not detect radio emission, setting 3$\sigma$ limits of 1.8 mJy at 150 MHz and 0.12 mJy at 400 MHz. We also discuss some of the possible reasons why no radio emission was detected from the HD 41004 binary system.

\end{abstract}

\begin{keywords}
Stars: planetary systems -- Stars: individual: HD~41004 -- Techniques: interferometric radio 
\end{keywords}

\section{Introduction}

In the last two decades, the exoplanet population has increased substantially. There have been several studies aimed at measuring and characterizing  the properties of exoplanets and their dependence on host star properties \citep[e.g.,][]{gonzalez97,cumming08,Perryman11,batalha13,gaidos13,marcy14,dressing15,mulders15small,winn15, Narang18,Mulders18, Shkolnik18,Winn18, Gayathri20}.  Still, one of the remaining and a very challenging task in exoplanet research is determining the interior structures and compositions of exoplanets, for which the detection of magnetic fields would provide important constraints. Radio emission is probably the only way to detect magnetic fields of exoplanets unambiguously \citep{Griessmeier15}.  It places not only strong constraints on the exoplanet magnetic fields but also provides important clues to their internal structure \citep[e.g.,][]{Sanchez-Lavega04,  Reiners10, Lazio19}. All planets in our solar system that possess a strong magnetic field also emit at radio wavelengths \citep[e.g.,][]{burke55, Bigg64,Zarka98, Zarka00}. Exoplanets with strong magnetic fields are also thought to emit strongly at radio wavelengths \citep[e.g.,][]{Zarka97, farrell99, Zarka01a,Lazio17,Zarka18b,Griessmeier18c}.

Planets in our solar system emit radio emission via the electron cyclotron maser instability (ECMI) mechanism \citep[e.g.,][]{ Wu79, Zarka97, Zarka98, Zarka00,  Treumann06,Louis19}. According to this mechanism, energetic electrons gyrate and accelerate along the magnetic field lines of planets (to which they are confined) and produce radio emission \citep{Melrose82, Dulk85}.  This emission is highly circularly polarized. The radio emission from exoplanets is also thought to follow a similar mechanism.

ECMI requires three components: (i) a magnetic field; (ii) a source of energetic (keV) electrons; and (iii) anisotropy in the electron distribution function. According to the Radiometric Bode's law, the radio power from the planet $P_{\mathrm{rad}}$, is directly proportional to the input power $P_{\mathrm{in}}$, supplied to the electrons \citep[e.g.,][]{Desch84,Zarka97,farrell99}. The primary mechanisms that can input power to the electrons and hence supply the energetic electrons for ECMI are \textit{Stellar Wind Kinetic Energy}, \textit{Stellar Wind Magnetic Energy}, \textit{Coronal Mass Ejection} (CME), and \textit{Unipolar interaction} \citep{Zarka01a,griessmeier07,Lazio17,Zarka18}.

There have been several studies aimed at detecting the radio emission from exoplanets  (see \citealt{Griessmeier17b} and \citealt{Lazio17} for  up to date reviews, along with \citealt{Bastian18}, \citealt{gorman18}, \citealt{Lynch18}, and \citealt{Matthew19}). However, none of them have been successful in directly detecting radio emission from exoplanets. Recently \cite{Vedantham20} have reported the detection of radio emission from an M dwarf star, GJ 1151. The detected radio emission from GJ 1151 is consistent with ECMI expected from a planet. \cite{Vedantham20} model the emission properties to show that the emission is consistent with the theoretical expectations for host star interaction with an Earth-size planet with a one-to-five-day orbital period. However, follow-up radial velocity studies have not been able to detect the planet but have placed upper limits of a few times the mass of the Earth \citep{Pope20}, which is consistent with their models for radio emission.

This paper presents our analysis of the radio observations of the binary system HD~41004 using the Giant Metrewave Radio Telescope (GMRT). HD 41004Bb is thought to be the best candidate to observe radio emission from an exoplanet with an estimated flux density of 910 mJy and with a maximum emission frequency of 140 MHz \citep{Griessmeier17b}. HD~41004Bb also has a period of only 32 hrs, making observations of its entire orbit possible. The large expected flux density at a low frequency of 150~MHz (close to the expected maximum cyclotron frequency), along with the short orbital period, makes HD~41004Bb an excellent candidate to observe with GMRT. We used legacy GMRT data (retrieved using the GMRT online archive) \footnote{naps.ncra.tifr.res.in/goa/data/search} along with observations with the upgraded GMRT (uGMRT) to produce the deepest images of the HD~41004 field at 150~MHz and 400~MHz. In Sect. 2, we describe the HD~41004 system. The observations and data analysis are described in Sect. 3. We present our results in Section 4 and discuss them in Section 5.

\section{Target}

HD~41004 is a $\sim1.6 \pm 0.8$ {Gyrs old} \citep{Saffe05} binary system located 41.5~pc \citep{Bailer18,Gaia18} away from Earth. The system is host to a K1 V primary (HD~41004A)  star, and an M2 V secondary star (HD~41004B) t{hat are separated by 0.5" (21~AU)}.  HD~41004A {hosts a Jupiter-like planet with a projected mass of} 2.5~$M_\mathrm{J}$ at 1.7~AU  from the host star \citep{Zucker04}. HD~41004B, on the other hand, {is host to a massive companion with a projected mass of} 18.4~$M_\mathrm{J}$, HD~41004Bb, at an orbital distance of 0.018~AU (orbital period = 32 hrs) \citep{Zucker04}.
Since these planets are detected only via the radial velocity method, the masses given above are lower limits M$\sin{i}$, and the true mass of the planets are likely to be higher. This might affect some of the estimates of maximum cyclotron frequency and expected flux density. Furthermore, the {projected} mass of HD~41004Bb is 18.4 $M_\mathrm{J}$, which is greater than~13~$M_\mathrm{J}$, the planet/brown dwarf boundary, which makes HD~41004Bb a low mass brown dwarf. However, the same radio emission mechanism of ECMI is thought to operate for brown dwarfs and exoplanets \citep[e.g.,][]{Antonova07, Hallinan07, Hallinan08}. 

\subsection{Emission frequency}

{The maximum cyclotron frequency for radio emission from the planet is a function of the surface magnetic field of the planet at the pole}
and is given as: 
\begin{equation}
\nu_\mathrm{c}~=~2.8~\mathrm{MHz}\;[B_\mathrm{P}/\mathrm{G}]
\end{equation}
where $B_\mathrm{P}$ is the {polar} magnetic field strength of the planet in Gauss. {We approximate the polar surface magnetic} field for an exoplanet as a function of planet mass $M_\mathrm{P}$, planet radius $R_\mathrm{P}$, and its rotation rate $\omega_\mathrm{P}$ (For further details see the Appendix~A):
\begin{equation}
  B_\mathrm{P} = 8.4~\mathrm{G} \; \bigg (\frac{\omega_\mathrm{P}}{\omega_\mathrm{J}}\bigg ) \bigg (\frac{M_\mathrm{P}}{M_\mathrm{J}}\bigg )\,  \, \bigg (\frac{R_\mathrm{J}}{R_\mathrm{P}}\bigg ) \, 
  \label{eq2a}
\end{equation}
where the planet mass, planet radius, and rotation rates are normalized with respect to that of Jupiter. This gives 

\begin{equation}
\nu_\mathrm{c}~=~2.8~\mathrm{MHz}\;[B_{\mathrm{P}}/\mathrm{G}]~=~ 23.5~\mathrm{MHz} \, \bigg (\frac{\omega_\mathrm{P}}{\omega_\mathrm{J}}\bigg ) \bigg (\frac{M_\mathrm{P}}{M_\mathrm{J}}\bigg ) \, \bigg (\frac{R_\mathrm{J}}{R_\mathrm{P}}\bigg ) \, 
\label{eq1} 
\end{equation}

\noindent

Based on its proximity to the host star, HD~41004Bb is {likely to be tidally} locked such that its rotation period and the orbital period are the same  \citep[see][for the time scale of tidal locking]{Gladman96}. Using the values for HD~41004Bb ($M_\mathrm{P}=18.4$~$M_\mathrm{J}$, period = 32 hrs and a radius $R_\mathrm{P}$=1.08~$R_\mathrm{J}$ using the mass-radius relationship from \cite{Chen17}) in equation~(\ref{eq1}), we find the maximum cyclotron frequency to be 125~MHz. With a bandwidth of emission $\Delta \nu\sim {\nu_\mathrm{c}}/2$ \citep{Lazio07, griessmeier07,grissermer11} the emission should be observable in band~2 (150~MHz) of GMRT. 

Equation~(\ref{eq2a}) used to estimate the magnetic field strength is based on the scaling relations derived from the planets in our solar system.  These scaling relations might not be applicable in the case of close-in planets. \cite{Yadav17} have argued that the magnetic fields of close-in hot-Jupiters could be $\sim$10 times that predicted using the scaling laws. Higher values of magnetic fields would result in the maximum cyclotron frequency of the emission going up. More conservatively, if $B_P$ is 2 to 4 times that predicted by equation~(\ref{eq2a}), that would correspond to maximum emission frequency ranging from 250-500 MHz. The emission then would fall within band~3 (250-500 MHz) of uGMRT. Moreover, a peak cyclotron frequency of 400 MHz also corresponds to the case of the planet not being tidally locked and having a rotation rate similar to that of Jupiter (10 hrs).

HD~41004Ab, on the other hand, is at 1.7 AU from the host star and has a {M$\sin{i}$} of 2.5 $M_\mathrm{J}$. Given that the {projected mass} of HD~41004Ab  is within a  few factors of  Jupiter, we might expect that its magnetic field is comparable to that of Jupiter. Consequently, it would emit at frequencies below 100~MHz (see equation \ref{eq1}) and would not emit at frequencies detectable by the GMRT.

\subsection{Modeling the emission}
The predicted radio flux density, $S_{\mathrm{rad}}$, of an exoplanet is given as \citep{farrell99,Lazio04,griessmeier07}:
\begin{equation} 
S_{\mathrm{rad}}=\frac{P_{\mathrm{rad}}}{\Delta \nu \, \Omega \, d^2}
\label{eq2} 
\end{equation}
where $P_{\mathrm{rad}}$, is the radio power emitted by the exoplanet, $\Delta \nu $ is the bandwidth of emission, $\Omega$, the angle of the emission cone, and $d$ is the distance of the exoplanet from Earth.  Based on the studies of the radio emission from solar system planets, it has been shown that the radio power $P_{\mathrm{rad}}$, from the planets, is proportional to the incident/input power from the solar wind, i.e., $P_\mathrm{rad} = \eta P_\mathrm{in}$, where $P_\mathrm{in}$ is the input power and  $\eta$ is a proportionality constant which takes care of power that is deposited in the planets and the conversion efficiency of the deposited power to radio power.  The proportionality constant $\eta$ is derived by equating the input power for Jupiter (for the various input power models) with the observed average radio power from Jupiter during periods of high activity \citep{griessmeier07, Lynch18}.

There are several mechanisms by which the input power from the stellar wind can be supplied to the planet, as discussed in Section 1. Below we discuss the two prominent mechanisms for the input power \citep[e.g.,][]{griessmeier07, Lazio17, Lynch18, Zarka18}: 

\begin{figure}
\centering
\includegraphics[width=1\linewidth]{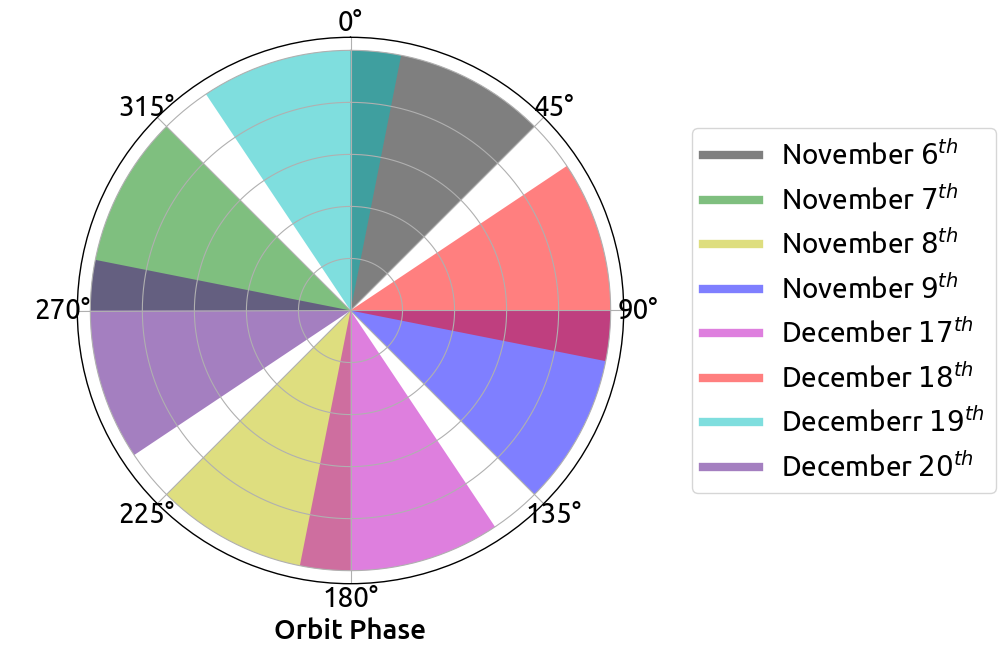}
\caption{The orbit phase coverage for HD~41004Bb during the GMRT observations at 150 MHz. We have assumed a phase of 0 at the beginning of the observation and then calculated the planet's phase at subsequent nights.  The different colors circle correspond to different nights.}
\label{fig2}
\end{figure}

\begin{enumerate}
\item \textit{Stellar Wind Kinetic Energy}: {The kinetic energy of the protons in the stellar wind impinging on the planet's magnetosphere provides the input power. The input power is proportional to the number density of the stellar wind, $n$, {at the location of the planet and the cube of the effective velocity of the stellar wind at the location of the planet, $v_{\mathrm{eff}}~=~\sqrt{v_{\star}^2 + v_k^2}$, where $v_{\star}$ is the stellar wind velocity at the location of the planet calculated using the Parker's isothermal solution \citep{Parker58,griessmeier07} and $v_k$ is the Keplerian velocity of the planet}.}

\begin{equation}
P_{\mathrm{in}}\propto n \, v_{\mathrm{eff}}^3
\label{eq5}
\end{equation}

\item \textit{Stellar Wind Magnetic Energy:} {The magnetic flux or the electromagnetic Poynting flux from the interplanetary magnetic field incident on the planet's magnetosphere provides the power input.  The input power, in this case, is proportional to the square of the strength of the interplanetary magnetic field perpendicular to the stellar wind flow, $ B_{\bot}$  at the location of the planet \citep[following][with the assumption that the stellar magnetic field strength is similar to that of the Sun]{Zarka07,griessmeier07} and effective velocity of the stellar wind, $v_{\mathrm{eff}}$}

\begin{equation}
P_{\mathrm{in}}\propto B_{\bot }^2 \, v_{\mathrm{eff}}
\label{eq6}
\end{equation}

\end{enumerate}

Recently, by studying the radio emission from the interaction between Jupiter and Ganymede and Jupiter and Io \cite{Zarka18} (also see \cite{Zarka17} and \cite{Griessmeier18c})  have shown that the emitted radio power is proportional to the Poynting flux from Jupiter's magnetosphere intercepted by Io or Ganymede. When combined with the scaling of radio power as a function of the incident magnetic power for planets in our solar system, this result indicates that the \textit{Stellar Wind Magnetic Energy} is the dominant mechanism that supplies the input energy for radio emission. However, since for the planets in our solar system, both the \textit{Stellar Wind Kinetic Energy} and the \textit{Stellar Wind Magnetic Energy} model give the same qualitative results \citep{Zarka07, Zarka18}, we have opted to investigate both models.  

{We model the flux density for HD~41004Bb at 125 MHz and 400 MHz using the two above mechanisms.} Taking the distance to the system from \cite{Bailer18} of 41.5 pc, $\Delta \nu$ = $\nu_c$/2 and $\Omega=1.6 \, sr$ \citep{Zarka04}  and making use of equations (\ref{eq2}), (\ref{eq5}) and (\ref{eq6}) we calculate the expected flux density from HD 41004Bb. For the maximum cyclotron frequency of 125 MHz, the expected flux density from the \textit{Stellar Wind Kinetic Energy} model is 3~mJy. At the same time, the expected flux density based on the \textit{Stellar Wind Magnetic Energy} model is 64.0~mJy.
{For the maximum cyclotron frequency of 400 MHz the predicted flux density from the \textit{Stellar Wind Kinetic Energy} is 2.0 mJy  and from the \textit{Stellar Wind Magnetic Energy} model the predicted {flux density} is 44 mJy.}

The  flux  density  estimates  from  our  models are  significantly  lower  than  that  from  \cite{Griessmeier17b} ($S_{rad}=910$ mJy with  maximum emission frequency $\nu_c$=140 MHz), who, following \cite{Griessmeier07b}  and  \cite{grissermer11},  estimated  the  flux  density  from  HD41004Bb  using  the Stellar Wind Magnetic Energy model.  \cite{Griessmeier17b} have has  use  of  several  empirical  relations  \citep[from, e.g., ][]{New80,GrieSmeier04,griessmeier07} that relate the age of the host star to the stellar wind velocity, the number density of protons in the stellar wind, stellar rotation rate, and magnetic field of the star for modeling the flux densities. Since individual age measurements of main sequence stars are difficult, and the HD 41004 system's age is very uncertain (1.6 $\pm$ 0.8 Gyrs), we do  not  use  these  relations  in  our  modeling.  However, if we also make similar assumptions as made by \cite{Griessmeier17b}, we get a similar value for the predicted flux density at with a maximum emission frequency of 125 MHz.

\section{Observation and Data Analysis }

The HD~41004 field was observed on multiple nights in November and December 2009, with the legacy GMRT. The field was observed at 150~MHz, with a bandwidth of 6~MHz. In order to ensure the full rotational phase coverage of HD~41004Bb, the field was observed in two observing blocks consisting of four successive nights in each block (2009 November $6^\mathrm{th}$, $7^\mathrm{th}$, $8^\mathrm{th}$, $9^\mathrm{th}$  and 2009 December $17^\mathrm{th}$, $18^\mathrm{th}$, $19^\mathrm{th}$,  $20^\mathrm{th}$), with $\sim$4 hours observing per night (see Figure \ref{fig2}).  

For each of the eight observations, the phase center was set at the position of HD~41004. The quasars 3C147 and 3C48 were used as the flux and band-pass calibrator. The band-pass and flux calibrators were observed twice, at the beginning and the end of the observations. We used 0616-349 as the phase calibrator. It was observed at frequent intervals during the observations in a loop of 27 min on the source, HD 41004, and 5 min on the phase calibrator, 0616-349. 

To reduce the data, the Source Peeling and Atmospheric Modeling (SPAM) pipeline {\citep{Intema09,  Intema14b, Intema14}} was used. SPAM is an automated pipeline for reducing legacy GMRT observations at 150, 235, 325, and 610~MHz. SPAM includes direction-dependent calibration, modeling, and imaging for correcting ionospheric dispersive delay. The pipeline first converts the raw data (LTA or  Long Term Accumulation format)  to precalibrated visibility sets and then converts these precalibrated visibility sets into Stokes I continuum images. 

\begin{figure*}
\centering
\includegraphics[width=0.37\linewidth]{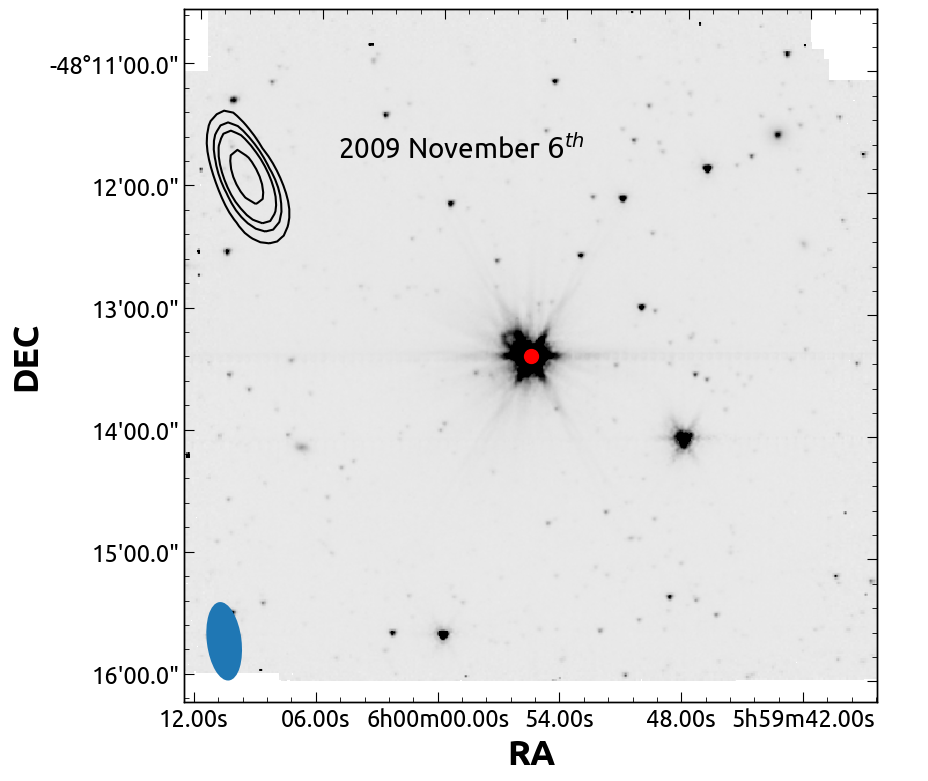}\includegraphics[width=0.37\linewidth]{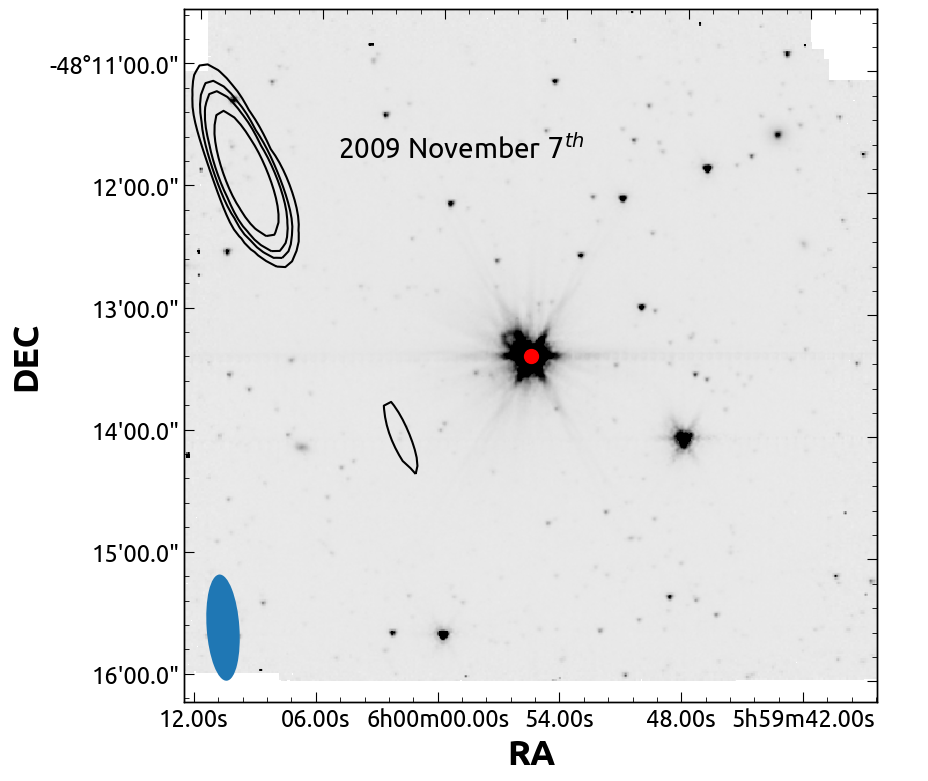}
\includegraphics[width=0.37\linewidth]{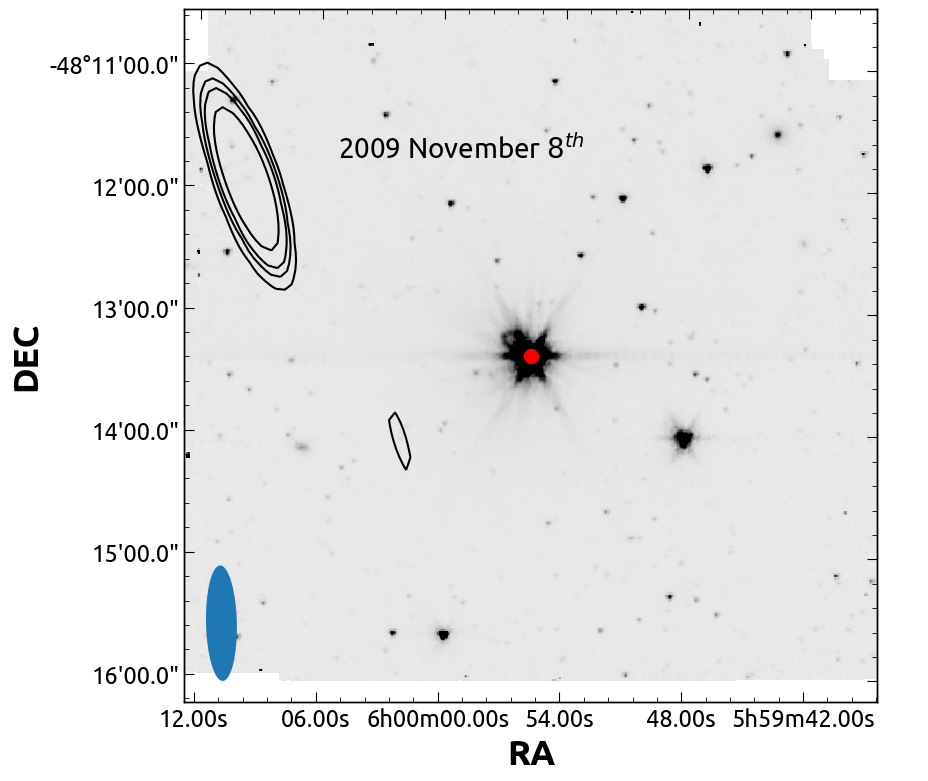}\includegraphics[width=0.37\linewidth]{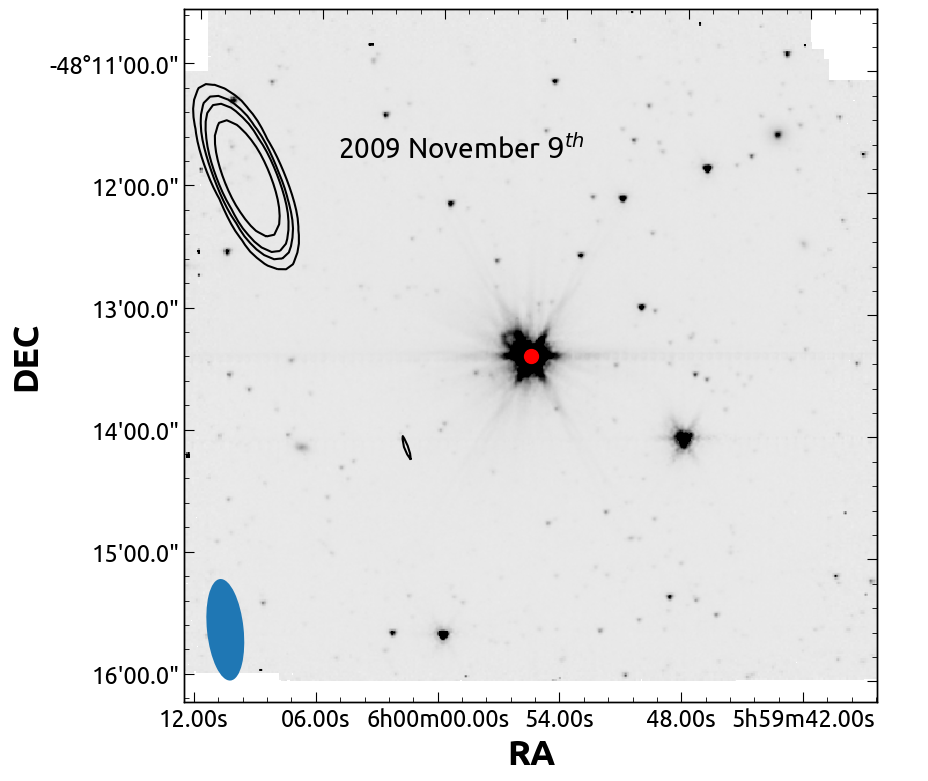}
\includegraphics[width=0.37\linewidth]{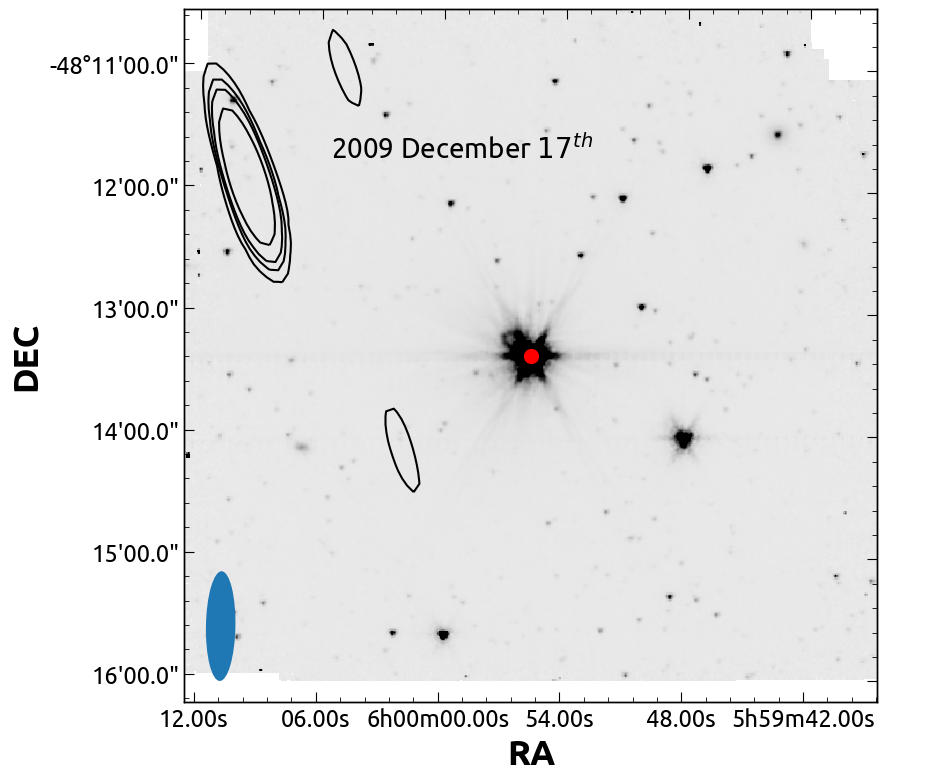}\includegraphics[width=0.37\linewidth]{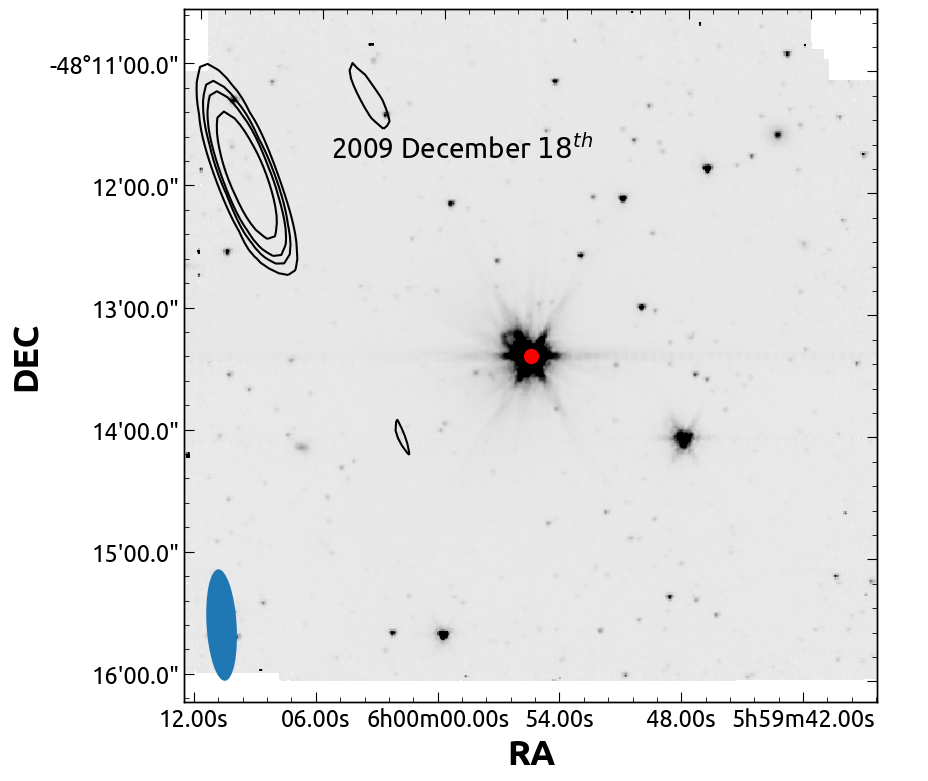}
\includegraphics[width=0.37\linewidth]{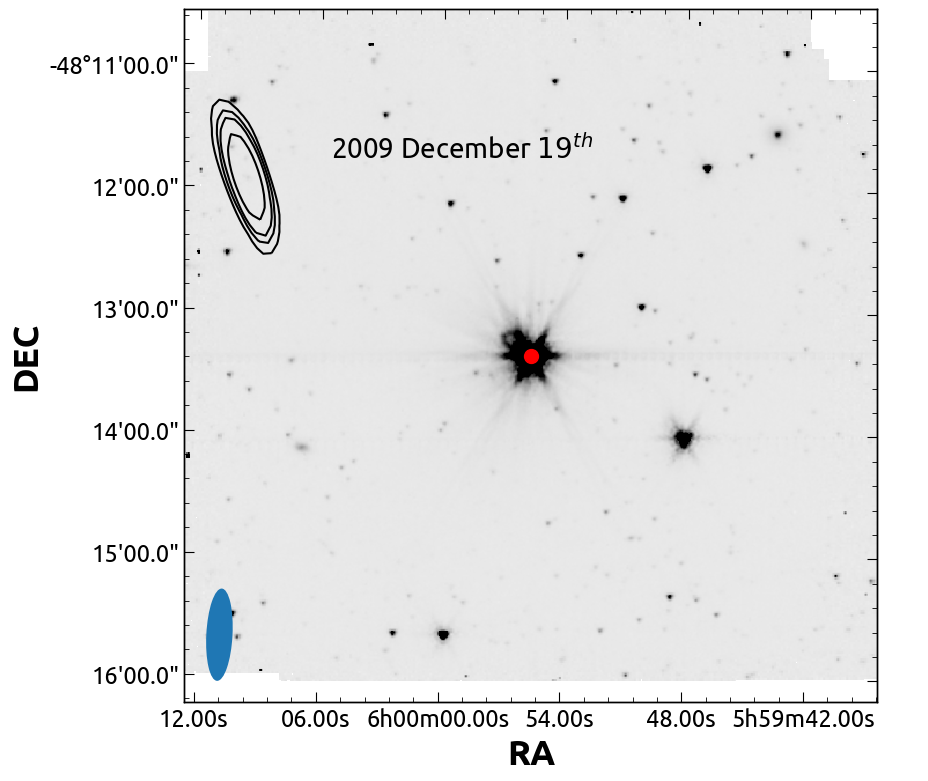}\includegraphics[width=0.37\linewidth]{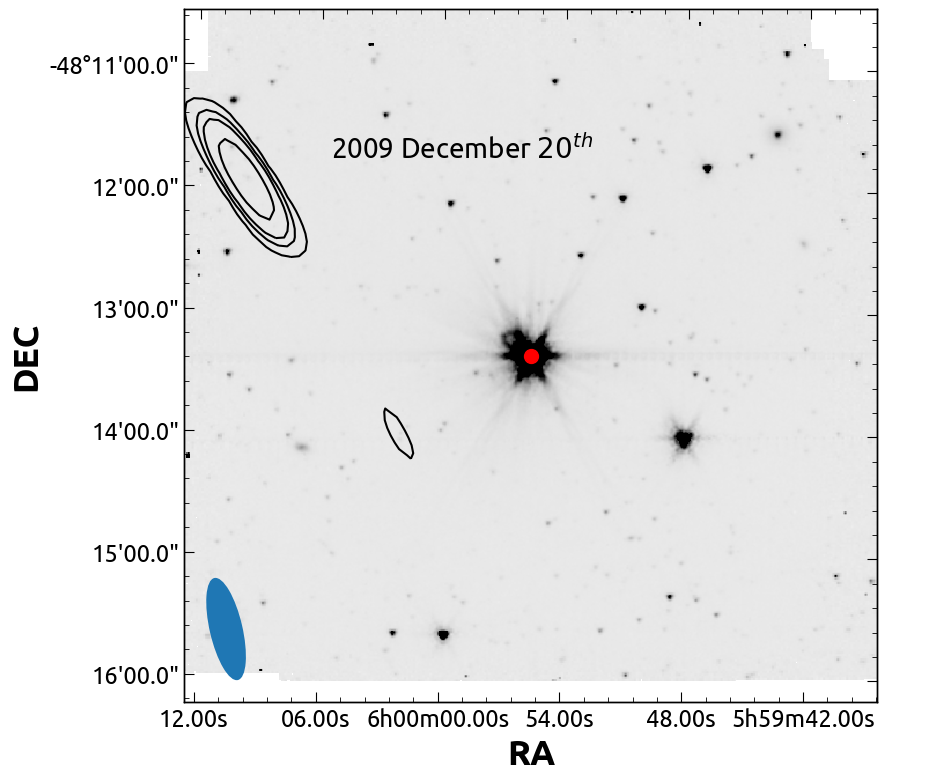}
\caption{The legacy GMRT image of the HD~41004 field at 150~MHz for each of the individual nights of observation  overlaid on the Spitzer IRAC~1~(3.6~$\mu m$) image of the HD~41004 field in greyscale. The red solid circle marks the position of the HD~41004 system. The contours plotted are 5, 10, 15, and 30~$\times\;\sigma$. The beam is shown as a blue ellipse at the bottom left corner. }
\label{fig2A}
\end{figure*}

We further carried out observations with the upgraded GMRT (uGMRT) in band~3 (250-500~MHz). The uGMRT observations of the HD~41004 field were made on 2018 October 30 and 31. The center frequency of the receiver was set at 400~MHz, with a bandwidth of 200~MHz. For each observation, the phase center was set at the position of HD~41004. 3C147 was observed as the primary flux density and band-pass calibrator twice during the observation, once at the beginning and once at the end. The phase calibrator used was 0616-349 and was observed for 5 minutes after 30 minutes on the target in a loop. 

The data were reduced using CASA with the uGMRT pipeline \citep{Ishwara20}. Initial flagging was carried out using the CASA task \textit{flagdata}. After flagging, the delay, band-pass, and gain calibration were performed.  The initial calibration was then removed, and the data were flagged again. After this, we re-calibrated the data. Then the imaging of the science target was carried out using the task \textit{tclean} in CASA. Four rounds of phase-only self-calibration with reducing solution intervals of 8, 4, 2, and 1 minute(s) were carried out. After convergence of phase-only self-calibration, five rounds of amplitude and phase self-cal with solution intervals of 8, 4, 2, 1, and 1 minute(s) were performed. From the first to last round of self-calibration cycle in both phase-only and amplitude and phase self-calibration, the number of clean iterations were doubled along with the lowering of clean threshold resulting in deeper images in subsequent cycles.  In between each self-calibration iteration, flagging based on residuals was performed, which improved image quality in each cycle (see \cite{Ishwara20} for more details). Afterward, a primary beam correction to the image was also applied using the {CASA task \textit{wbpbgmrt}}\footnote{https://github.com/ruta-k/uGMRTprimarybeam} to produce the final image.

\begin{figure}
\centering
\includegraphics[width=1\linewidth]{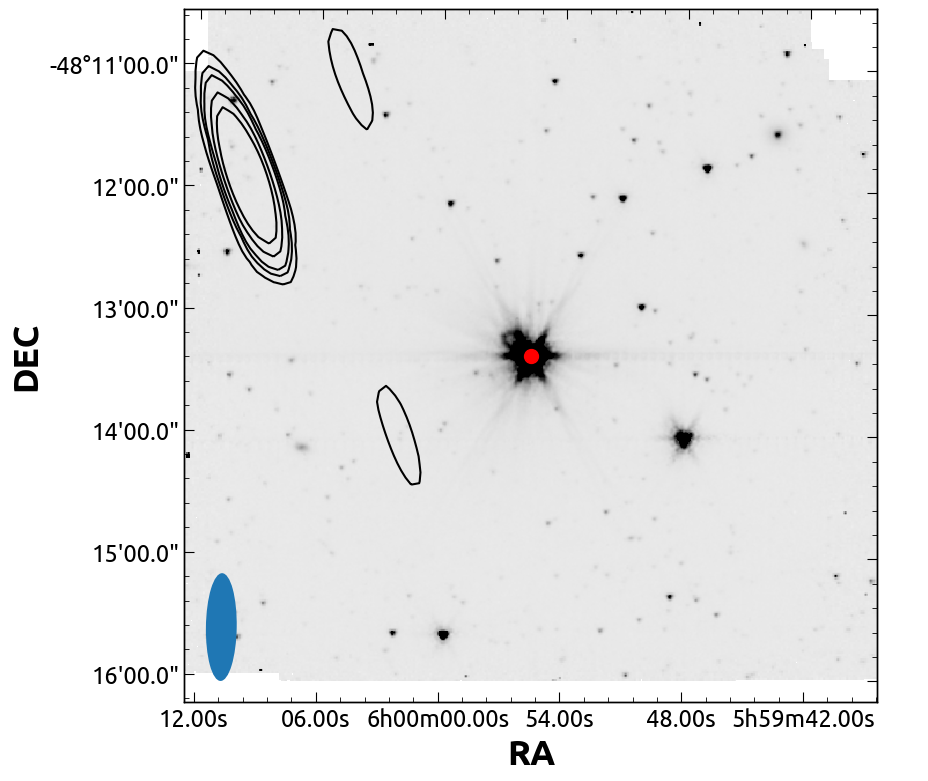}
\caption{The legacy GMRT image of the HD~41004 field at 150~MHz after combining the observations from  November $8^\mathrm{th}$, November $9^\mathrm{th}$, December $17^\mathrm{th}$, December $18^\mathrm{th}$, and December $19^\mathrm{th}$ overlaid on the Spitzer IRAC~1~(3.6~$\mu m$) image of the HD~41004 field in greyscale.  The solid red circle marks the position of the HD~41004 system. The contours plotted are 5, 10, 15, 30, 50~$\times\;\sigma$ with $\sigma=0.6$ mJy. The beam is shown as a blue ellipse at the bottom left corner. At this scale, we cannot resolve HD~41004A and HD~41004B.} 
\label{fig3}
\end{figure}

\section{Results}
No radio emission was detected from either HD~41004Ab or HD~41004Bb at 150~MHz or 400~MHz. The HD~41004 field was observed at 150~MHz with the legacy GMRT for eight nights with $\sim$4 hrs of observations per night, totaling to 32 hrs of observation time.  No emission was detected for each of the $\sim$4 hr observations. The rms values achieved for each of the eight nights are listed in Table \ref{table1}, and the images are shown in Figure \ref{fig2A}. To make a deeper image, we combined the eight observations. We were able to achieve an rms of 0.8~mJy after 32 hrs of observations. However, if we exclude the observations from 2009 November $6^\mathrm{th}$, 2009 November $7^\mathrm{th}$, and 2009 December $20^\mathrm{th}$, which have rms higher than the median rms for the eight nights, we produce an image with an rms of 0.6~mJy. This is shown in Figure \ref{fig3}. This rms is similar to the lowest rms achieved with GMRT for an exoplanet field at 150~MHz \citep{Hallinan13} (also see Figure \ref{fig5}).

\begin{figure}
\centering
\includegraphics[width=1\linewidth]{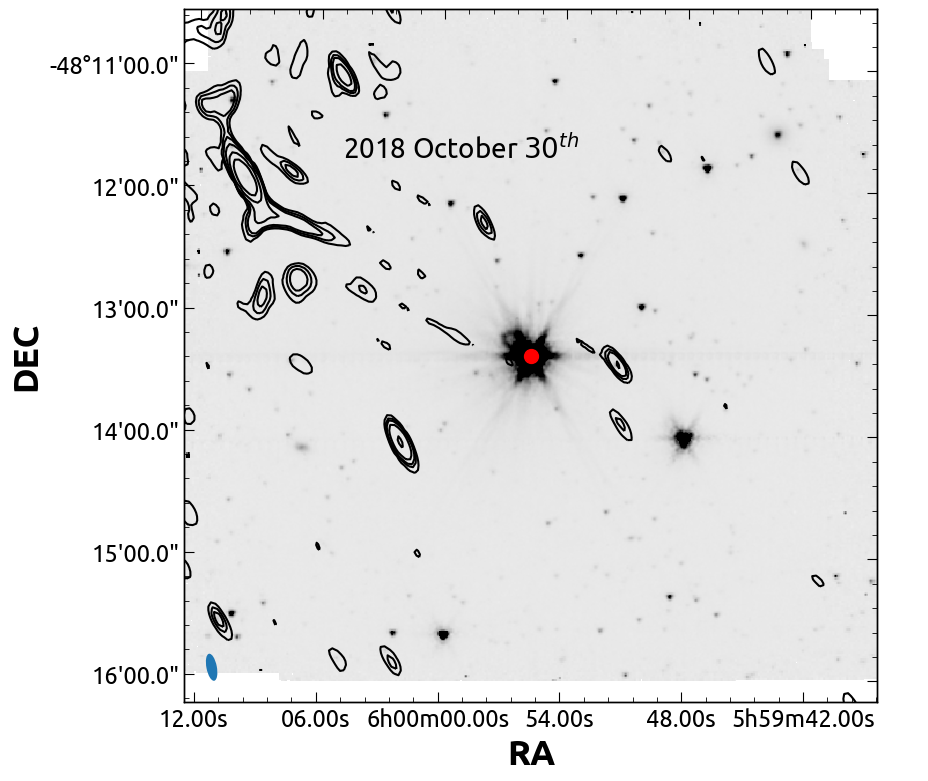}
\includegraphics[width=1\linewidth]{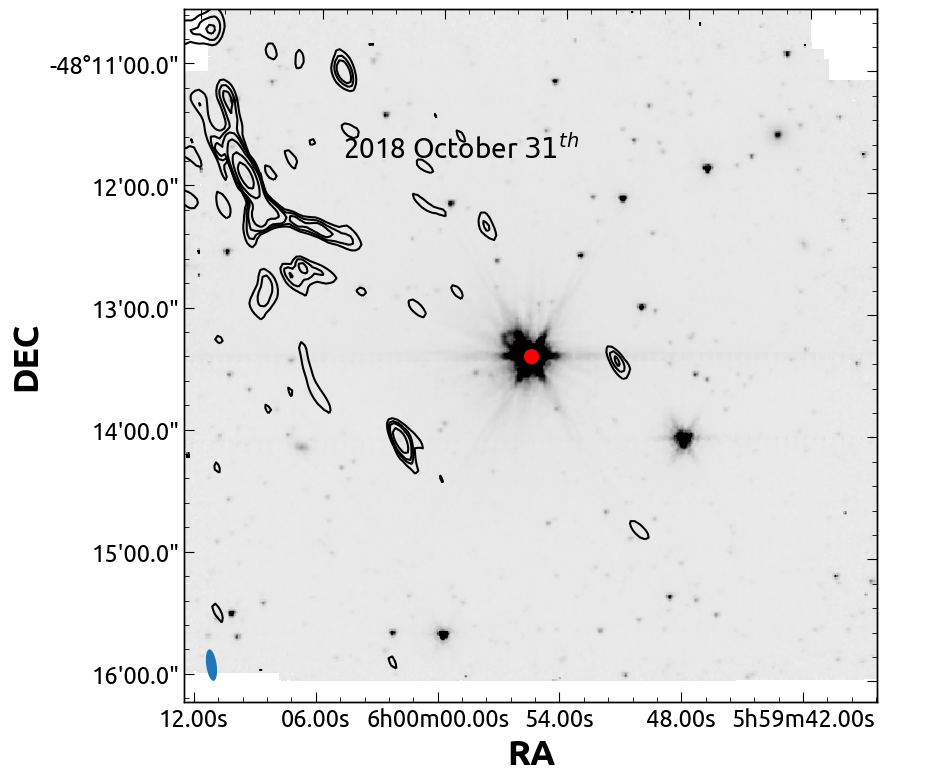}
\caption{The radio image of the HD~41004 field at 400~MHz for two different nights 2018 October 30  (a) and 2018 October 31 (b). These images are overlaid on the Spitzer IRAC 1 image of the HD~41004 field in grey scale. The red solid circle marks the position of HD~41004 system. The contours plotted are 5, 10, 15, 30, 100 and 300$\times \sigma$ where $\sigma=40~\mu$Jy for the  night of October 30 and $\sigma=70~\mu$Jy for the night of October 31. The beam is shown as a blue ellipse at the bottom left corner. } 
\label{fig4}
\end{figure}

We also observed the HD~41004 system in band~3 (250-500~MHz) with uGMRT at 400 MHz with a bandwidth of 200 MHz. An advantage of observing in band~3 {(250-500~MHz)} with uGMRT is that it offers better sensitivity than that can be obtained in band~2 (150~MHz) of GMRT  \citep{YG}. No radio emission was detected from the HD~41004 systems on either of the nights. From our uGMRT observations, we obtained an rms of 40~$\mu$Jy on the night of October 30th and an rms of 70~$\mu$Jy on October 31st at 400~MHz after 3 hrs of observations on each night. The final images are shown in Figure \ref{fig4}(a) and Figure \ref{fig4}(b). 

Our observations reached sensitivities that could have easily resulted in detection, yet no emission was detected toward HD 41004Bb. At the maximum emission frequency of 150 MHz, the \textit{Stellar Wind Magnetic Energy} model predicts a flux density of 64 mJy, while the achieved $3\sigma$ sensitivity is 1.8 mJy. Similarly, at the maximum emission frequency of 400 MHz, the predicted flux density (based on the \textit{Stellar Wind Magnetic Energy} model) is 44 mJy, while the $3\sigma$ sensitivity achieved was 0.12 mJy (also see Table \ref{Table2}).

Using the observed flux density upper limits, we can place limits on the radio power emitted by the HD 41004Bb system. These calculations are based on the assumption that there is a steady-state quiescent component to the emission and that the Earth was in the emission cone of the planet during the observations. Using Equation~ \ref{eq2} and assuming the solid angle of the beamed emission, $\Omega=1.6 \,sr$ \citep{Zarka04} and $\Delta \nu = \mathrm{observed \, frequency}/2$, the upper limit to radio power at 150 MHz  is $3.5 \times 10^{15}$~W. At 400~MHz, the upper limit to radio power is $6\times 10^{14}$~W. These upper limits are about 3 to 4 orders higher than the average power during periods of high activity from Jupiter $P_{\mathrm{rad}}=2.1 \times 10^{11}$~W \citep{Zarka04}.

\begin{table*}
\centering
\begin{tabular}{|c|c|c|c|c|c|}
\hline
 &  &  &  &  &  \\
Band & Date & Central Frequency & Bandwidth & Observation time & Image RMS \\
&---  & (MHz) & (MHz) & (hrs) &~(mJy/beam) \\ 
 &  &  &  &  &  \\\hline
 &  &  &  &  &  \\
2 & 2009 November $6^\mathrm{th}$ & 150 & 6 & 4 & 2.1 \\
 &  &  &  &  &  \\
2 & 2009 November $7^\mathrm{th}$ & 150 & 6 & 4 & 1.2 \\
 &  &  &  &  &  \\
2 & 2009 November $8^\mathrm{th}$  & 150 & 6 & 4 & 0.9 \\
 &  &  &  &  &  \\
2 & 2009 November $9^\mathrm{th}$  & 150 & 6 & 4 & 1.0 \\
 &  &  &  &  &  \\
2 & 2009 December $17^\mathrm{th}$ & 150 & 6 & 4 & 1.1 \\
 &  &  &  &  &  \\
2 & 2009 December $18^\mathrm{th}$ & 150 & 6 & 4 & 1.1 \\
 &  &  &  &  &  \\
2 & 2009 December $19^\mathrm{th}$  & 150 & 6 & 4 & 1.0 \\
 &  &  &  &  &  \\
2 & 2009 December $20^\mathrm{th}$  & 150 & 6 & 4 & 1.7 \\
 &  &  &  &  &  \\
2 & Combining all & 150 & 6 & 32 & 0.8 \\
 &  &  &  &  &  \\
2 & Combining observations from & 150 & 6 & 20 & 0.6 \\
& 2009  November $8^\mathrm{th}$, November $9^\mathrm{th}$, 2009 December $17^\mathrm{th}$, &  & &  &\\
&December $18^\mathrm{th}$, and December $19^\mathrm{th}$  &  & &  &  \\
 &  &  &  &  &  \\
\hline
 &  &  &  &  &  \\
3 & 2018 October $30^\mathrm{th}$ & 400 & 200 & 3 & 0.04 \\
 &  &  &  &  &  \\
3 & 2018 October $31^\mathrm{th}$ & 400 & 200 & 3 & 0.07\\
 &  &  &  &  &  \\
\hline
\end{tabular}
\caption{Summary of the observations and the combined images.}
\label{table1}
\end{table*}

\section{Discussion}
\begin{figure*}
\centering 
\includegraphics[width=1\linewidth]{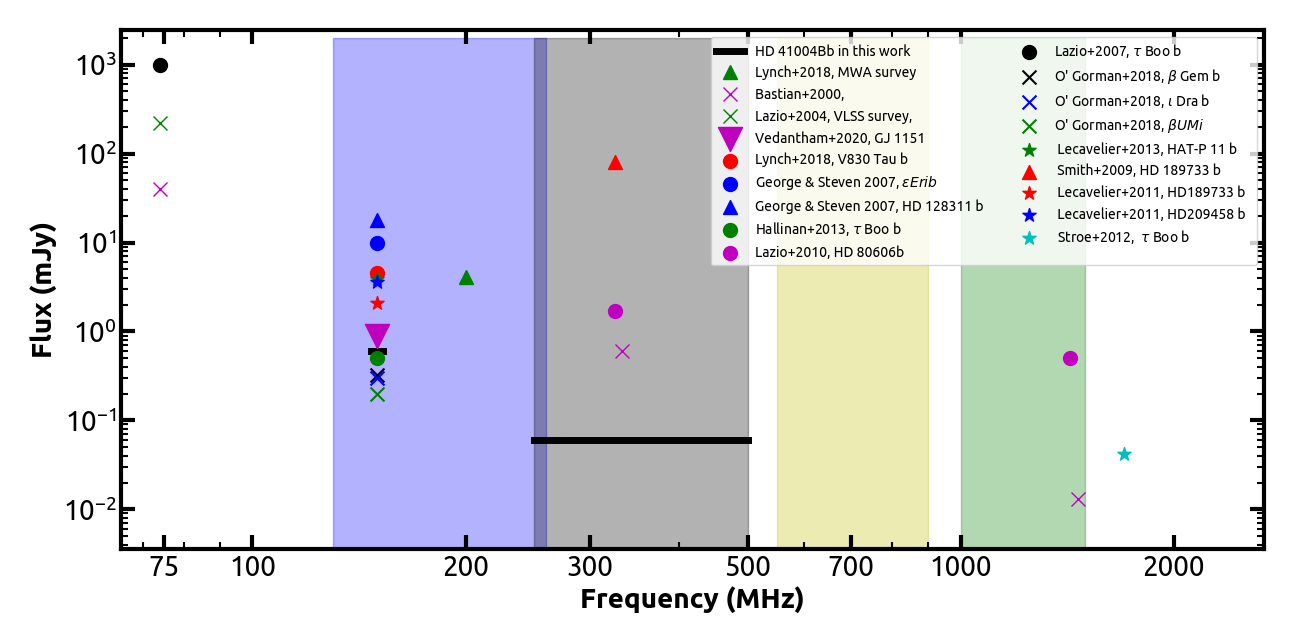}
\caption{The rms  flux densities {for} some of the previous attempts at detecting radio emission from exoplanets as a function of the frequency of observations \citep{Lazio04,Lynch18,Bastian00,Lazio07,George07,Hallinan13,gorman18,Etangs13,Etangs11,Lazio10,smith09,Stroe12} along with the rms sensitivity we reached in this work (black solid line) for the 150MHz and 400 MHz uGMRT observations. {Also shown is the detection of radio emission from the system GJ 1151 by \citet{Vedantham20}}. The uGMRT bands are shown in different colors ({blue band~2 (120-250 MHz), grey band~3 (250-500 MHz), yellow band~4 (550-900 MHz), and green band~5 (1000-1450 MHz).}}
\label{fig5}
\end{figure*}

We have made some of the deepest observations at 150~MHz and 400~MHz of an exoplanet field.  In Figure \ref{fig5}, we compare rms sensitivities of observations that have been carried out to detect radio emission from exoplanets.  {Also shown is the detection of radio emission from the system GJ 1151 by \cite{Vedantham20} as a magenta downwards triangle.} {The flux density from GJ~1151 (0.89 mJy) is comparable to the rms flux density (0.6 mJy) we achieved at band-2 (150~MHz)}.

In the following subsection, we discuss possible reasons why no radio emission was detected from HD~41004Bb and how our deep observations allow us to place critical constraints on various radio emission mechanisms.

\subsection{Overestimation of flux  and saturation of emission}

One of the reasons {why} we did not detect any radio emission could be that the radio emission is too weak at these frequencies, and {the} models of the emission mechanisms might need further revisions. To calculate the expected flux density, several assumptions, and scaling relationships had to be used (to calculate, e.g., the stellar wind proton density, the stellar wind speed, stellar magnetic field). These assumptions and scaling relationships are based on the Sun. If these relationships and assumptions do not hold for an M type star, {the,} modeled flux densities might be incorrect.

Further, \cite{Jardine08} suggested that close-in planets might lie inside the magnetosphere of the star. Due to this, the magnetospheres of these close-in planets are determined by the magnetic pressure of the stellar magnetosphere and not by the ram pressure of the stellar wind. As the planetary orbital distance decreases, the electron density $\rho_e$ rises, so the number of available electrons increases. However, this is exactly balanced by the shrinking of the planet's magnetosphere, which reduces the cross-section of the planet available for radio emission \citep{Jardine08}. This may result in the saturation of the output power, even while the input power increases.

In hot-Jupiter, the magnetospheric convection may saturate, and, hence, they may not dissipate the total incident's magnetic energy from the stellar wind  \citep{Nichols16}.  Due to this, the emission from these hot-Jupiters {may} not follow the  Radiometric Bode's Law, and the expected flux densities would be overestimated.

If this is the case, the observed flux density upper limits that we derive in Section 4 can be used to constrain by how much the constant of proportionality $\eta$ (that relates input power to radio power) for HD 41004Bb differs from that of Jupiter. The proportionality constant, $\eta$, used in calculating the expected flux densities, was obtained by comparing the estimated input power to Jupiter (for the two models) with the observed radio power from Jupiter. However, $\eta$ could be different for HD~41004Bb. An upper limit to the ratio of  $\eta_{HD}$ (constant of proportionality for HD~41004Bb) and $\eta_J$ (constant of proportionality for Jupiter)  can be calculated as a multiplicative factor  $\kappa_\mathrm{max}$, by dividing the observed $3\sigma$ flux density upper limits, $S_{3\sigma}$ with the modeled flux density, $S_{\mathrm{rad}}$ such that:

\begin{equation}
    \frac{\eta_{HD}}{\eta_{J}} = \kappa_\mathrm{max} = \frac{S_{3\sigma}}{S_{\mathrm{rad}}}
    \label{eq7}
\end{equation}

\noindent
In Table \ref{Table2} we summarise the values of the predicted flux density for the two models (\textit{Stellar Wind Kinetic Energy} and \textit{Stellar Wind Magnetic Energy}) for the two peak emission frequencies that we have considered along with the 3$\sigma$ rms sensitivity achieved and the values of  {$\kappa_\mathrm{max}$}. The value of {$\kappa_\mathrm{max}$}  is between 0.003 to 0.6 for the different models/hypotheses we have considered in this study. 

\subsection{Incorrect maximum cyclotron frequency estimates}
For HD~41004Bb, the maximum cyclotron frequency, assuming tidal locking, is 125~MHz. Recently, however, \cite{Yadav17} have argued that due to the inflation of hot-Jupiters, their magnetic fields can be much stronger than that predicted using scaling laws that do not take the heating of these planets into account. Stronger magnetic fields imply higher frequency than predicted using empirical relations previously used in various studies. Furthermore, if the planet is weakly magnetized, the emission would be at a much lower frequency.

\begin{table*}
\begin{tabular}{|c|c|c|c|c|}
\hline &  &  &   &  \\
Model & Maximum cyclotron frequency & Modeled flux density & achieved 3$\sigma$ rms & {$\kappa_\mathrm{max}$} \\
-- & MHz & mJy & mJy & -- \\  &  &    &  &  \\
\hline
&  &    &  &  \\
Stellar Wind Kinetic Energy & 125 & 3 & 1.8 & 0.6 \\  &  &    &  &  \\
Stellar  Wind  Magnetic  Energy  & 125 & 64 & 1.8 & 0.03 \\  &  &    &  &  \\
Stellar Wind Kinetic Energy & 400 & 2 & 0.12 & 0.06 \\  &  &  &  &  \\
Stellar  Wind  Magnetic  Energy & 400 & 44 & 0.12 & 0.003 \\  &  &    &  &  \\
\hline
\end{tabular}
\caption{Summary of the predicted flux density and the 3$\sigma$ rms sensitives achieved during our observations. Also shown is the value of   {$\kappa_\mathrm{max}$} which is the upper limit to the ratio of $\eta_{HD}$ to that of $\eta_J$ (see equation \ref{eq7}) }
\label{Table2}
\end{table*}

\subsection{Variable/Episodic emission}
The low-frequency radio emission from the solar system planets is highly variable over time. For example, Jupiter's decameter emission varies on timescales of several minutes and can produce emissions that are ten times higher than the median level \citep{Zucker04, gorman18}.  The radio emission from exoplanets could also be episodic. Episodic radio bursts have been seen form isolated brown dwarfs \citep[e.g.][]{Berger01,Hallinan08,Wolszczan14,Kao16,Williams17}. The emission from GJ~1151 is also variable \citep{Vedantham20}, and hence long term monitoring of exoplanet hosts would be required to rule out the variable/episodic nature of emission.

\subsection{Beamed Emission}
The other possible reason for non-detection could be that the Earth was outside the planet's emission beam. Cyclotron maser emission is known to be confined within a beam. The decameter emission of Jupiter is only detectable over certain rotational phase ranges because the emission is narrowly beamed  \citep{gorman18, Hallinan13}.  Assuming that the magnetic axis is aligned with the rotation axis and that emission is produced near the planetary poles, the typical beaming angle is between 50$^\circ$ $-\,$ 60$^\circ$ {\citep{Melrose82, Dulk85, Zarka98, Treumann06}.} Such beamed emission will only be observable from Earth during a particular phase of the orbit of the exoplanet; our near-complete orbital coverage for HD 41004Bb at 150~MHz partially mitigates this problem. An alternate approach to detect beamed emission is to observe several sources so that in a few cases, the emission is beamed along the line of sight.

\subsection{Quenching of emission}
If the local plasma frequency, $\nu_{p}$ is too high, the cyclotron emission cannot propagate.  For a plasma distribution with density $n$ (in $\mathrm{cm}^{-3}$), the plasma frequency, $ \nu_{p}$ is given as 

\begin{equation}
    \nu_{p}=\sqrt{\frac{n\,e^2}{\pi m_e}}\sim 8.98\,\mathrm{kHz} \sqrt{n}
\end{equation}

\noindent
The plasma will absorb emission at frequencies lower than the $ \nu_{p}$. {For radio emission to be detected, the {maximum cyclotron frequency}}, {$\nu_{c}$} within the source region, must be greater than the $ \nu_{p}$ at every point along the line of sight such that $ \nu_{c}>> \nu_{p}$.

If the above condition is not satisfied, then the radio emission from the exoplanet would be quenched due to the local plasma \citep{griessmeier07,hess11}. The number density of electrons $n$ from the stellar wind, at the position of HD~41004Bb {(following \citep{GrieSmeier04,  Griessmeier07b, Lynch18})} taking the stellar mass, age, spectral type as well as the wind velocity and the orbital distance of the planet into account is $\sim 6.6\times 10^4$~$\mathrm{cm}^{-3}$, that gives $ \nu_\mathrm{P}$~=~2.2~MHz, much below the emission frequency.

{Further}, for close-in hot-Jupiters/ brown dwarfs, the upper layers of the atmosphere are heated by the stellar radiation, leading to an inflated and extended upper atmosphere. The outer layers also become highly ionized, which increases the electron density, resulting in a high plasma frequency. If this frequency exceeds the {maximum cyclotron frequency}, then the emission will be quenched (also see  {\citep{Daley17,Daley18, Weber17, Weber17b, Weber18}).}

\section{Conclusion }
In this paper, we carried out one of the most sensitive radio observations of the HD~41004 field with GMRT at 150MHz and 400~MHz. We do not detect radio emission from the {HD~41004 system}.  However, we can place a very strong and robust $3\sigma$ upper limits of 1.8~mJy at 150~MHz and 0.12 mJy at 400~MHz on the {steady-state or quiescence component of the} radio emission (provided that the emission is beamed towards us). These are some of the lowest limits achieved at these wavelengths and are much lower than the theoretically predicted {flux densities}. We have provided some plausible reasons for the non-detection of radio emission from this system.

\section{Acknowledgement}
{We thank the referee for their insightful comments and suggestions that have resulted in a significant improvement of the manuscript.}
This work is based on observations made with the Giant Metrewave Radio Telescope, which is operated by the National Centre for Radio Astrophysics of the Tata Institute of Fundamental Research and is located at Khodad, Maharashtra, India. We thank the GMRT staff for efficient support to these observations. We acknowledge the support of the Department of Atomic Energy, Government of India, under the project  12-R$\&$D-TFR-5.02-0700. M.T. is supported by MEXT/JSPS KAKENHI grant Nos. 18H05442, 15H02063, and 22000005. Part of this research was carried out at the Jet Propulsion Laboratory, California Institute of Technology, under a contract with the National Aeronautics and Space Administration. This research has made use of the NASA Exoplanet Archive, which is operated by the California Institute of Technology, under contract with the National Aeronautics and Space Administration under the Exoplanet Exploration Program. This research has also made use of NASA's Astrophysics Data System Abstract Service and the SIMBAD database, operated at CDS, Strasbourg, France.

\section{Data availability}

The data presented in this article are available on the GMRT archive at https://naps.ncra.tifr.res.in/goa/, 
and can be accessed with proposal ids $17\_066$ and $35\_001$.

\appendix

\section{Magnetic field of planets}

For solar system planets it has been suggested that the magnetic moment of a planet is proportional to its angular momentum \citep{farrell99}.
\begin{equation}
\mu_\mathrm{P} \propto L_\mathrm{P} \sim \omega_\mathrm{P}\, M_\mathrm{P} \, R_\mathrm{P}^{2}
\end{equation}

\noindent
where $L_\mathrm{p}$ is the angular momentum of the planet, $\mu_\mathrm{P}$ is the magnetic moment of the planet and $R_\mathrm{P}$, $M_\mathrm{P}$ and $\omega_\mathrm{P}$ are the radius, mass and rotation rate of the planet. For Jupiter, 
\begin{equation}
    \mu_\mathrm{J} \sim \omega_\mathrm{J}\, M_\mathrm{J} \, R_\mathrm{J}^{2} 
\end{equation}
and also,
\begin{equation}
    \mu_\mathrm{J}=~4.2\,[\mathrm{G}] \, R_\mathrm{J}^{3} \, 
\end{equation}
from \cite{Connerney81}.

\vspace*{0.25cm}

\noindent
Dividing equation~(A1) by equation~(A2) and substituting the values from equation~(A3)

\begin{equation}
{\mu_\mathrm{P}}  = \bigg (\frac{\omega_\mathrm{P}}{\omega_\mathrm{J}}\bigg ) \bigg (\frac{M_\mathrm{P}}{M_\mathrm{J}}\bigg ) \, \bigg (\frac{R_\mathrm{P}}{R_\mathrm{J}}\bigg )^{2}   4.2 [\mathrm{G}] \,  R_\mathrm{J}^{3} \, 
\end{equation}

\vspace*{0.25cm}
\noindent
The magnetic field $B_\mathrm{P}$ at the poles is given as: \begin{equation}
    B_\mathrm{P}~=~\frac{2\,\mu_\mathrm{P}}{R_\mathrm{P}^3}~=~8.4 [G] \, \bigg (\frac{\omega_\mathrm{P}}{\omega_\mathrm{J}}\bigg ) \bigg (\frac{M_\mathrm{P}}{M_\mathrm{J}}\bigg )\,  \, \bigg (\frac{R_\mathrm{J}}{R_\mathrm{P}}\bigg )
\end{equation}

\noindent
and thus, we obtain
\begin{equation}
    \nu_\mathrm{c}~=~2.8~\mathrm{MHz}\;[B_{\mathrm{P}}/\mathrm{G}]~=~ 23.5~\mathrm{MHz} \, \bigg (\frac{\omega_\mathrm{P}}{\omega_\mathrm{J}}\bigg ) \bigg (\frac{M_\mathrm{P}}{M_\mathrm{J}}\bigg ) \, \bigg (\frac{R_\mathrm{J}}{R_\mathrm{P}}\bigg ) \,
\end{equation}

\bibliographystyle{mnras}
\bibliography{ms}
\bsp    
\label{lastpage}
\end{document}